   \documentstyle[agupp]{article}

\lefthead{SUBRAMANIAN ET AL.}

\righthead{RELATIONSHIP OF CMES TO STREAMERS}

\cpright{AGU}{1996}

\authoraddr{Prasad Subramanian, K.~P. Dere, N.~B. Rich
and R.~A. Howard, Code 7660, Naval Research Laboratory, 
4555 Overlook Ave SW, Washington, DC 20375, USA (e-mail: 
psubrama@mahanadi.nrl.navy.mil; dere@halcyon.nrl.navy.mil; 
nathan@calliope.nrl.navy.mil; howard@cronus.nrl.navy.mil)}

\slugcomment{To appear in the {\it Journal 
of Geophysical Research}, 1999.}

\makeatletter
\let\reset@font\empty
\makeatother

\begin{document}

%
%

\title{The Relationship of Coronal Mass Ejections \\
to Streamers}


%
%

\author{Prasad Subramanian}
\affil{Center For Earth Observing and Space Research, George
Mason University, Fairfax, VA 22030, USA.}

\author{K.~P. Dere}
\affil{Code 7660, Naval Research Laboratory,
Washington, DC 20375, USA.}

\author{N.~B. Rich}
\affil{Interferometrics, Inc., Chantilly, VA 22021, USA.}

\author{R.~ A. Howard}
\affil{Code 7660, Naval Research Laboratory,
Washington, DC 20375, USA.}

\begin{abstract}
We have examined images from the Large Angle Spectroscopic Coronagraph (LASCO) to study the relationship of Coronal Mass 
Ejections (CMEs) to coronal streamers.
We wish to test the suggestion (Low 1996) that CMEs arise from flux ropes
embedded in a streamer erupting, thus disrupting the streamer. 
The data span a period of two years near sunspot minimum through a period of increased activity as sunspot numbers increased. We have used LASCO data from the C2 coronagraph which records Thomson scattered white light from coronal electrons at heights
between 1.5 and 6R$_\odot$. Maps of the coronal streamers have been constructed from LASCO C2 observations at a height of 2.5R$_\odot$ at the east and west limbs. We have superposed the corresponding positions of CMEs observed with the C2 coronagraph onto the synoptic maps. 
We identified the
different kinds of signatures CMEs leave on the streamer structure at this
height (2.5R$_\odot$). We find four types of CMEs with respect to their effect on streamers
\begin{enumerate}
\item
CMEs that disrupt the streamer
\item
CMEs that have no effect on the streamer, even though they are related to it. 
\item
CMEs that create streamer-like structures
\item
CMEs that are latitudinally displaced from the streamer.
\end{enumerate} 
This is the most extensive observational study 
of the relation between CMEs and streamers to date.
Previous studies using SMM data have made the general statement that CMEs are mostly
associated with streamers, and that they frequently disrupt it.
However, we find that approximately 35\% of the observed CMEs bear no
relation to the pre-existing streamer, 
while 46\% have no effect on the observed streamer, 
even though they appear to be related to it. Our conclusions thus differ
considerably from those of previous studies.
\end{abstract}

\begin{article}
\section{Introduction}

 Streamers are the most prominent  
observable features of
the large-scale magnetic field of the sun.
CMEs, in turn, are thought to arise from the destabilization of
large scale, closed magnetic structures/arcades associated with streamers. 
The effect of CMEs on streamers is therefore crucial to building a 
complete picture of CME initiation and subsequent propagation.
The data from LASCO observations
comprises the most detailed set of white light observations
of CMEs to date.
We use synoptic white light maps obtained from images taken
by the C2 coronagraph of the LASCO instrument to map the streamers. 
We then superimpose CME locations
inferred from C2 images on these synoptic maps. 
This dataset reveals several new aspects of the association
between CMEs and streamers that have not been appreciated in previous
studies such as those by Hundhausen (1993).
In particular, it reveals that a substantial fraction of CMEs have very little
effect on the existing streamer structure, and that many of them do not seem to bear any relation to a streamer.
A relatively smaller percentage of CMEs also seem to \em create \em 
coronal structures that can be referred to as stalks or legs.
We seek to elaborate on such aspects in this paper, and place
them in the context of the current state of knowledge about the overall
issue of CME initiation and propagation.

We start by 
examining previous observational work on the relationship between CMEs and streamers, and on the magnetic topology of the post-CME corona in \S 2. 
We next examine the theoretical motivations for investigating
the effect of Coronal Mass ejections on streamers in \S 3.
We move on to discussing the details of the method we use
to construct white light synoptic maps, and the manner in which
we mark the locations of CMEs on them in \S 4. \S 5 presents the salient 
results that can be drawn from our analysis of the data. A discussion of
the significance of these results in the context of our current state of
understanding about CMEs is contained in \S 6.


   %
   %
   %

\section{Observational Motivations}\label{text}

\markcite{{\it Howard et al.} [1985]}
published a comprehensive catalog of CMEs observed with the Solwind instrument.
They coined the phrase ``streamer blowout'' to denote events where the streamer
that existed prior to the CME would start swelling and brightening, and
eventually disappear as a result of the CME. They found that such
events were typically slow in comparison to other kinds of CMEs. 
\markcite{{\it Illing \& Hundhausen} [1986]} subsequently discussed a
filament eruption and CME observed by the SMM coronagraph that resulted
in a streamer blowout. \markcite{{\it Hundhausen} [1993]} published a comprehensive review of CMEs observed by the SMM spacecraft.
He concluded that an appreciable fraction of CMEs observed during 1984
were ``bugle'' CMEs, which were so named because the pre-CME streamers
 looked like
bugles facing to the left on a conventional white light synoptic map. 
The bugle-like
structures on the synoptic maps are a consequence of the fact that the 
streamer widens prior to being blown out by the CME.
He also concluded that CMEs are often associated with 
disruptions of the streamer structure. This, together with the idea that streamers are manifestations
of the large-scale coronal magnetic field, is used to support the thesis that
CMEs are associated with the destabilization of large-scale magnetic
structures. The generally accepted notion that CMEs are associated with
the streamer belt is also attributed to \markcite{{\it Hundhausen} [1993]}. 
The theme of CMEs being associated with large-scale,
closed magnetic structures in the corona has also been emphasized by
\markcite{{\it Bravo et al.} [1998]}. \markcite{{\it McAllister and
Hundhausen} [1996]} conclude that 73 \% of coronal arcade events observed with the YOHKOH Soft X-Ray telescope are associated with streamer belts. This, together with the
generally accepted notion that X-ray arcade events are proxies for CMEs, reinforces the notion that CMEs are very often associated with streamer belts.
The present study however sketches a more complex picture of the association of CMEs with streamers.

Our study differs from that of \markcite{{\it Hundhausen} [1993]} in the
following aspects:
\begin{itemize}
\item
Our conclusions are based on examination of data from observations spanning
two years, whereas Hundhausen's conclusions about the relationship of
CMEs to streamers are based on data spanning one year (see Figure 1).
\item
The superior sensitivity of the LASCO instrument allows us to arrive at
more quantitative and definite conclusions regarding the effect of
CMEs on streamers as compared to the study of 
\markcite{{\it Hundhausen} [1993]}.
\end{itemize}

\section{Theoretical Motivations}
We first proceed to undertake a brief review of streamer models.
Early streamer models were motivated primarily by eclipse 
observations. A comprehensive review of such models is given in
\markcite{{\it Koutchmy and Livshits} [1992]}. In particular,
the early model of \markcite{{\it Pneuman and Kopp} [1971]} modeled the
streamer as a magnetostatic structure containing an axisymmetric current
sheet. Subsequent treatments have attempted to include the effects of
the solar wind on the streamer structure in a self-consistent manner.
\markcite{{\it Wang et al.} [1997]} published a streamer model that was
motivated by LASCO observations. They interpreted the observed streamers as arising from Thomson-scattering electrons that are concentrated around
a warped current sheet encircling the sun. Their
model employs a potential extrapolation of the photospheric magnetic
field up to a source surface of approximately 2-3 R$_\odot$. While it 
reproduces the gross features of the streamer belt near solar minimum rather well, it does not include free currents in the corona, and consequently
does not address any dynamical phenomena like the effect of CMEs on the 
streamer.

We next briefly review models of CME initiation and propagation.
Our intent in doing so will be to identify model predictions
that address the issue of the effect of CMEs on streamers.

Some models of CME initiation rely on the picture of a flux rope breaking through/disrupting an overlying magnetic field 
arcade, resulting in the observable
CME in the field of view of a coronagraph. 
These are based on models published by \markcite{{\it Low} [1996]} which 
suggest that the cavity underlying helmet streamers is a flux rope
containing prominence material. There exist equilibrium solutions where the magnetic flux rope is attached to the Sun, and also in the case where it is
detached from the Sun. 
The concept of a detached flux rope embedded in a helmet streamer has been
taken as a starting point for a number of numerical simulations. 
\markcite{{\it Guo et al.} [1996]} simulate the 
dynamic response
of the streamer to the emergence of a current-carrying magnetic bubble.
If the magnetic field carried by the bubble is oppositely directed to that of the overlying streamer, they find that reconnection occurs at the 
flanks of the emerged structure where the current density is maximum. 
Depending upon the initial energy in the current-carrying magnetic bubble, 
they find that the overall streamer structure either 
remains quasi-static, or disrupts. 
\markcite{{\it Wu and Guo} [1997]} also address the role of magnetic 
buoyancy in disrupting
the streamer. They find that low-density flux ropes disrupt the
streamer faster than high-density ones. The overall conclusion
seems to be that all models of flux rope-streamer systems disrupt the
streamer during a CME; the differences in the densities of the flux 
rope manifest
themselves in the speed at which the streamer is disrupted.

Some other groups (see, for instance, \markcite{{\it Mikic and Linker} [1997]}, 
\markcite{{\it Steinolfson} [1994]})
model CMEs by considering reconnection processes at the
bottom of the magnetic arcade that results in the formation of
a pinched-off plasmoid. The formation and subsequent propagation
of the plasmoid can be interpreted as a disruption of the streamer.
\markcite{{\it Antiochos} [1998]}, on the other
hand, argues that reconnection occurs \em above \em the arcade that
actually erupts, so as to allow the erupting structure to rise.

\section{Procedure}
\subsection{The data}
The data used in this study was obtained by the Large Angle Spectroscopic 
Coronagraph (LASCO) instrument aboard the Solar and Heliospheric Observatory
(SOHO). A detailed description of LASCO is given in \markcite{{\it 
Brueckner et al.} [1995]}.
Briefly, LASCO consists of 3 coronagraphs, C1, C2 and C3. Here, we have
used images from the C2 coronagraph which cover the corona from a distance
of 1.5 out to 6 R$_\odot$, and are obtained with a typical cadence of
30 to 60s between images. The LASCO images record all of the CMEs that
occur during periods of operation, as well as the evolution of the coronal
streamers. The CME times and locations with respect to the coronal streamers
are the basis for this study. The procedure is described in the following
sections. We have used the data from the start of LASCO operations in
January 1996 until the period in June 1998 when SOHO was lost for 2.5 months.
This covers times of very low solar activity near solar minimum through a
period of increasing activity as the sun enters the new activity cycle. Figure 1 displays the monthly and smoothed sunspot number index (obtained from the
Sunspot Index Data Center at http://www.oma.be/KSB-ORB/SDIC). We have indicated the period covered by the LASCO observations and contrasted it with that covered by the SMM observations (\markcite{{\it Hundhausen} [1993]}) on the figure.

\subsection{Construction of White Light Synoptic Maps of the Corona}
The white light synoptic maps are constructed from intensities measured 
at a height of 2.5 solar radii over the east and
west limbs of images from the C2 coronagraph (Figures 2 and 3).
Following the convention for such synoptic maps, time increases from
the right to the left.
The start date corresponds to the time when the central meridian is on the east (west) limb; hence, the start date of a west limb map is the same date as the middle of the east limb map (half a rotation as the sun rotates east to west). 
A more complete set of synoptic maps for the whole SOHO/LASCO mission are available at http://lasco-www.nrl.navy.mil/carr$_{-}$maps.
All the maps display the ratio of the observed signal to a background model to account for scattered light and
the dust (F) corona.
Vertical black lines are missing data blocks.
Synoptic maps for Carrington rotation 1919 (February 1997) onward 
have a longitudinal resolution of 0.5 degrees (0.91 hours) per pixel; 
the prior rotations 
 have a longitudinal resolution of 1 degree (1.82 hours) per pixel.
All the maps have a latitudinal resolution of 1 degree per pixel. 
\subsection{Synoptic Locations of CMEs}
We mark the footpoints of the CME when it emerges into the field of
view of the C2 coronagraph at 2.2 R$_\odot$. 
The latitudinal extent of the CME is measured in the plane of the sky.
The longitude of the CME is taken to be the longitude of the east/west limb
at the time of observation. No attempt is made to account for the possible
longitudinal displacement of the CME out of the plane of the sky.
The coordinates of the footpoints of the CME are converted into
heliographic coordinates and superimposed on the white light synoptic
maps. 
The `X's in Figures 2 and 3 are the counterparts of
CME footpoints as observed in individual images, and 
represent the latitudinal extrema of the
CME. In Figures 2 and 3, a pair of `X's connected by a vertical dashed
line appearing against the background of a white light 
synoptic map thus represents a CME.
Halo CMEs, which were first reported by \markcite{{\it Howard et al.} [1982]}, 
are interpreted as
CMEs moving towards (or away from) the earth. 
Halo events usually originate around 90 degrees from the limb. 
They are typically wide, sometimes spanning an entire solar diameter, so that their footpoint locations do not fit the general scheme used here.
We mark the halo events with a vertical string of `H's, 
and don't join them with a dashed line, as we do with the other CMEs.
We include the halo events in our catalog for the sake of completeness.

\section{Results}
Our observations span a period between January 1996 and June 1998, which ranges from near minimum solar activity to progressively increasing activity, as shown in Figure 1.
An examination of the data (examples of which are shown in Figures 2 and 3) reveals that the CMEs
 can be placed into
four categories, based on their relationship to the streamer.
We describe below the classifications, and the percentage of
CMEs that fall into each of the categories. We have recorded a 
total of 375 CMEs during our observation period, of which 30 are halo CMEs.
The percentage figures mentioned in each category are based upon an 
examination of the entire dataset covering the period of observations 
mentioned above.
Owing to space constraints, we do not show the entire
dataset in this paper. We display  the datasets corresponding to Carrington
rotations 1925 and 1932 in Figures 2 and 3 respectively. These two
figures contain examples of all the four categories of events described 
below.

\begin{enumerate}
\item
\em Creates streamer \em : 
This category includes events where considerably more coronal material is present at the location of the CME after its eruption compared to that prior to the eruption. It includes the creation of post-CME structures that can be variously described as stalks, streaks, legs and so on. Figures 4 a-c show a 
representative example of an event where a `leg' is created following
a CME. The leg is created towards the northern boundary of the CME, and
persists for approximately 1.5 days following the CME.
This event, which occurs on 1997/07/24, can be discerned on the east limb
synoptic map of Figure 2.
This class of events
relates most closely to those discussed by \markcite{{\it Kahler and Hundhausen} [1992]}.
We have placed 29 events in this category, which corresponds to 
{\bf 8.4\%} of the total number of non-halo CMEs recorded.

\item
\em Displaced from streamer \em:
This category includes events that are displaced from whatever streamer
structure is present.
Figure 5 shows an example of such an event. This event
occurs on 1997/07/24, and appears on the west limb synoptic map of Figure 2.
 93 events fall into this category, which corresponds
to {\bf 27\%} of the total number of non-Halo events recorded. Events in this category do not have any effect on the streamer.

\item
\em No effect \em :
This category includes events which, unlike
those included in the previous category, do straddle the existing streamer 
structure, but seem to have no effect on the streamer. 
There are several examples of such events on the synoptic maps in
Figures 2 and 3. These CMEs are typically dim, and are evident only
in running difference images. This implies that the CMEs are probably
displaced from the plane of the sky.
We record 160 events in this category, which corresponds to
{\bf 46\%} of all the non-Halo events.

\item
 \em Streamer blowout \em : 
In this category, 
we include all events where part, or all of the streamer
structure that exists prior to the CME disappears (or becomes significantly
reduced in intensity) following the CME. Insofar as the effect of CMEs
on streamers is concerned, this category of events is the most dramatic, and 
has been most widely 
discussed in the literature (\markcite{{\it Howard et al.} [1985]}, \markcite{{\it Hundhausen} [1993]}).
Furthermore, as noted in \S 3, most theoretical models of CME initiation
and propagation seem to predict a disruption of the streamer of some kind.
Figures 6 a-b show an example of a streamer blowout event. This event occurs on 1998/02/04, and appears on the west limb synoptic map of Figure 3.
We have recorded 56 events in this category, which corresponds to
{\bf 16\%} of the total number of non-Halo events recorded.
\end{enumerate}

The creation of streamer-like structures following a CME bears no relation
to the streamer structure that existed prior to the CME. We may therefore
add the numbers for categories 1 (Creates Streamer) and 2 (Displaced from Streamer), and conclude that 35\%
of the non-Halo CMEs observed bear no relation to the pre-existing
streamer. Similarly, events in categories 3 (No effect on Streamer) and 4 (Streamer Blowout) are associated with 
the streamer. We therefore add the numbers ascribed to these two
categories and conclude that 63\% of all the non-halo CMEs we observe
are related to the streamer. Figure 7 summarizes the statistics presented
in this section.

\section{Discussion}
The relationship of CMEs to streamers is an area which has been examined
only by \markcite{{\it Hundhausen} [1993]} and  
\markcite{{\it McAllister and Hundhausen} [1996]} so far.
The results obtained from our study, which surveys the relationship between
CMEs and streamers for the period Jan 1996 - Jun 1998, modify some of 
the prevalent perceptions on this subject.
Some statements in the literature
(e.g., Hundhausen, 1993) suggest that CMEs originate 
from the streamer belt, and that they
often disrupt the streamer. 
\markcite{{\it McAllister and Hundhausen} [1996]} find that 73\% of
X-ray arcade events are associated with streamers. Since X-ray arcade
events observed by YOHKOH are generally considered to be proxies for
CMEs, this tends to suggest that a large fraction of CMEs are associated
with streamers. We find that about 63\% of the
observed CMEs are associated with streamers. Thus,
our conclusions are fairly consistent with those of \markcite{{\it McAllister and Hundhausen} [1996]}. 
 \markcite{{\it Hundhausen} [1993]} concludes that
approximately 50\% of the CMEs observed in 1984 result
in streamer disruptions.
However, we find that only a small percentage (16\%)
of CMEs result in disruption of the streamer.

The most surprising
result of our analysis, in our view, is the large number
of CMEs in category
3 of the preceding section. We observe that a significant 
fraction (46\%) of the observed 
CMEs do not have any effect on the streamer, although they seem to be
associated with it.
We now discuss how such events pose a paradox in the context
of the flux rope model for CMEs.
In the flux rope model of \markcite{{\em Low} [1996]}, for instance, the flux rope is located at the base of the streamer. If ejected, this should disrupt the streamer.
There have been a number of papers in the recent literature that report
ejections of helical magnetic flux ropes
(\markcite{{\it Dere et al.} [1998]}, \markcite{{\it Chen et al.} [1997]}),
as suggested by \markcite{{\em Low} [1996]}. There have also
been a number of theoretical treatments that envisage a CME as a magnetic
flux rope, as outlined in \S 3. 
One possible explanation for events which overlap the streamer, but do not seem to affect it in any
way is as follows: the CME could be longitudinally 
displaced from the limb, and its two-dimensional projection in the
plane of the sky could therefore be rather faint.
The observed streamer, on the other hand, is in the plane of the sky, and
is therefore bright. 
However, the longitudinal extent of a typical
flux rope is expected be comparable to, if not
larger than that of
 undulations in the warped current sheet that are manifested as the streamer.
Therefore, even if an ejected flux rope is longitudinally displaced from the
observed streamer structure, it is difficult to reconcile it with the
fact that it has no effect whatsoever on the streamer.

Another view of the CMEs classified in category 3 is as follows.
Figure 8 shows the relative contributions of matter at different
angles out of the plane of the sky to the total integrated brightness.
It is evident from Figure 8 that a structure situated 45 degrees 
away from the plane of
the sky would seem 20\% as bright as one situated in the plane of the
sky. The CMEs in category 3 are typically dim, which implies that they
are displaced from the plane of the sky. If we assume that these CMEs are
about 20\% as bright as the brightest ones observed, we can roughly
estimate from Figure 8 that they are displaced by at least 45 degrees from
the plane of the sky, and perhaps not spatially related to the streamers
observed in the synoptic maps. Such projection effects could therefore be
one possible explanation for the paradox posed by events in category 3, which
overlap the streamer, but do not disrupt it.

\section{Conclusions}
We have drawn the following conclusions from our study of CMEs over the
period spanning January 1996 -- June 1998:
\begin{enumerate}
\item
8.5\% of all the CMEs observed create streamer-like structures after they
erupt.
\item
27\% of all the CMEs are latitudinally displaced from the streamer.
They thus seem to bear no relation to the streamer.
\item
46\% of all the CMEs overlap the streamer, but seem to have no effect on it.
This large fraction is especially puzzling if most CMEs are flux ropes 
embedded in the streamer, for they would be expected to disrupt the streamer
when they erupt.
\item
Only 16\% of all the observed CMEs disrupt the streamer.
\end{enumerate}
These conclusions considerably modify the prevalent perceptions that CMEs
are usually associated with streamers, and that they frequently disrupt it.
The large percentage of CMEs that do not affect the streamer despite being
associated with it is especially intriguing. One possible explanation for 
such CMEs is that they are longitudinally displaced from the plane of
the sky, as explained in \S 6. Even so, it casts doubt on the theme
of CMEs being associated with disruptions of 
large scale, closed structures in the corona.

%
%

\acknowledgments

The SOHO/LASCO data used here are produced by a consortium of the Naval Research Laboratory (USA), Max-Planck-Institut fuer Aeronomie
(Germany)), Laboratoire d'Astronomie (France), and the University of Birmingham (UK). SOHO is a project of international cooperation between ESA
and NASA. This work was made possible with funding from NASA.

\end{article}

\newpage

\begin{figure}
\caption{This figure displays the monthly sunspot number and
a smoothed fit to the data from 1980 to 2000. We have marked the period
covered by the SMM observations reported by Hundhausen (1993) and that
covered by the LASCO observations.}
\end{figure}

\begin{figure}
\caption{Figures 2 and 3 are examples from the dataset we have used in this work. Figure 2a shows the east limb white light synoptic map for Carrington rotation 1925, with CME footpoints superimposed on it. Figure 2b shows the corresponding map for the west limb. The streamer is the bright structure along the equator. 
The `X's in Figures 2 and 3 are the counterparts of
CME footpoints as observed in individual images, and 
represent the latitudinal extrema of the
CME. In Figures 2 and 3, a pair of `X's connected by a vertical dashed
line appearing against the background of a white light 
synoptic map thus represents a CME.
It is seen from the synoptic maps of Figures 2 and 3 that CMEs affect the
streamer in different ways. Individual images for the specific CMEs
marked on Figure 2 are shown in Figures 4 and 5.}
\end{figure}

\begin{figure}
\caption{Figure 3 shows the east and west limb white light snyoptic maps for Carrington rotation 1932, with CME footpoints superimposed on it. 
Individual images for the specific CME marked on Figure 3b are shown in 
Figure 6.}
\end{figure}

\begin{figure}
\caption{Figures 4a, 4b and 4c show a sequence of images taken with the LASCO C2 coronagraph. They show a CME taking off on the north-eastern limb. It creates a bright stalk that was not present prior to the CME, and persists for approximately 1.8 days following the CME. This CME is an example of an event in category 1, \S 5, ``Creates Streamer''. This CME occurred on 07/24/97, and is
marked on the synoptic map of Figure 2a. It is clearly evident from
Figure 2a that a bright, long-lived stalk is created after the passage of the
CME.}
\end{figure}

\begin{figure}
\caption{This figure shows a CME occurring on the north-western limb. It is an example of a CME that is displaced from the main streamer structure, as described in category 2, \S 5, ``Displaced from Streamer''. This event occurred
on 07/24/97, and is marked on the synoptic map of Figure 2b. As is evident
from Figure 2b, the CME is displaced northward from the streamer.}
\end{figure}

\begin{figure}
\caption{Figures 6a and 6b show a CME which disrupts and ``blows out'' the streamer. It is an example of the class of events described in category 4, \S 5, ``Streamer Blowout''. This event occurred on 02/04/98, and is marked on the synoptic map of Figure 3b. It can be clearly seen from Figure 3b that this CME
disrupts the streamer.}
\end{figure}

\begin{figure}
\caption{Figure 7a embodies the central result of our paper. It shows the relative number of events in each of the categories described in \S 5. The events in categories 1 and 2 bear no relation to the pre-existing streamer. We add the percentages assigned to these two categories and conclude that 35\% of all the CMEs are unrelated to the pre-existing streamer, as shown in figure 7b. Similarly, adding the numbers assigned to categories 3 and 4 leads to the conclusion that 63\% of CMEs are related to the pre-existing streamer, as shown in figure 7b.}
\end{figure}

\begin{figure}
\caption{Figure 8 gives the relative contribution of electrons along the 
line of sight to the total brightness.  For example, electrons
located more than 40 degrees from the plane of the sky, contribute
only about 20\% of the total signal.}
\end{figure}

\end{document}